# Observation of $S=1/2$ quasi-one-dimensional magnetic and magneto-dielectric behavior in a cubic $SrCuTe_2O_6$

B. Koteswararao,[1,2,3] S. K. Panda,[4] R. Kumar,[5] Kyongjun Yoo,[2] A. V. Mahajan,[5] I. Dasgupta,[4,6] B. H. Chen,[3] Kee Hoon Kim,[2,*] and F. C. Chou[3,+]

[1]*School of Physics, University of Hyderabad, Central University PO, Hyderabad 500046, India*

[2]*CeNSCMR, Department of Physics and Astronomy, and institute of applied physics, Seoul National University, Seoul 151-747, South Korea*

[3]*Center of Condensed Matter Sciences, National Taiwan University, Taipei 10617, Taiwan*

[4]*Centre for Advanced Materials, Indian Association for the Cultivation of Science, Jadavpur, Kolkata-700032, India*

[5]*Department of Physics, Indian Institute of Technology Bombay, Mumbai 400076, India*

[6]*Department of Solid State Physics, Indian Association for the Cultivation of Science, Jadavpur, Kolkata 700 032, India*

**Abstract**

We investigate magnetic, thermal, and dielectric properties of $SrCuTe_2O_6$, which is isostructural to $PbCuTe_2O_6$, a recently found, Cu-based 3D frustrated magnet with a corner-sharing triangular spin network having dominant first and second nearest neighbor (*nn*) couplings [B. Koteswararao, *et al.* Phys. Rev. B **90**, 035141 (2014)]. Although $SrCuTe_2O_6$ has a structurally similar spin network, but the magnetic data exhibit the characteristic features of a typical quasi-one-dimensional magnet, which mainly resulted from the magnetically dominant third *nn* coupling, uniform chains. The magnetic properties of this system are studied via magnetization ($M$), heat capacity ($C_p$), dielectric constant ($\varepsilon'$), measurements along with *ab-initio* band structure calculations. Magnetic susceptibility $\chi(T)$ data show a broad maximum at 32 K and the system orders at low temperatures $T_{N1} \approx 5.5$ K and $T_{N2} \approx 4.5$ K, respectively. The analysis of $\chi(T)$ data gives an intra-chain coupling, $J_3/k_B$, to be about $\approx -42$ K with non-negligible frustrated inter-chain couplings ($J_1/k_B$ and $J_2/k_B$). The hopping parameters obtained from LDA band structure calculations also suggest the presence of coupled uniform chains. The observation of simultaneous anomalies in $\varepsilon'(T)$ at $T_{N1}$ and $T_{N2}$ suggests the presence of magneto-dielectric effect in $SrCuTe_2O_6$. A magnetic phase diagram is also built based on $M$, $C_p$, and $\varepsilon'$ results.

## I. INTRODUCTION

Magnetism in one-dimensional (1D) antiferromagnetic (AFM) $S=1/2$ systems has been very interesting due to their inherent tendency to uphold the strong quantum fluctuations, which lead to a ground state (GS) with quasi-long range order (LRO) [1]. In a 1D chain, the involvement of the frustrated next nearest neighbor (*nnn*) coupling (*J′*) to its nearest neighbor (*nn*) coupling (*J*) exhibits a varieties of exotic ground states ranging from spin disordered, gapped states to spin ordered states with simultaneous ferroelectric order. The emerging ground states depend upon the type as well the strength of those couplings (*J* and *J′*). For instance, in case of $S=1/2$ AFM chain, when the ratio of *nnn* to *nn* coupling, (*J″/J*) is about 0.5, the resultant GS opens up a spin gap to its first excited state [2]. On the other hand, the involvement of *nnn* AFM coupling to a ferromagnetic (FM) uniform chain also causes the competition between these two interactions and induces some nontrivial magnetic orderings like spiral, helical, etc. Theoretically, the helical magnetic structure has been proposed for │*J″/J*│ > 0.25 [3,4].

Experimentally, such a state has already been realized for many $S=1/2$ chain systems, including $LiCu_2O_2$, $LiCuVO_4$, $Li_2ZrCuO_4$, $CuCl_2$, $CuBr_2$, $CuCrO_4$, *etc* [5-12]. These are few well-known examples of ribbon chains made by $CuX_4$ plaquette (where X=O, Cl, and Br) and possess nontrivial spin structures. Also, ferroelectricity was evidenced for most of the above mentioned systems, except $Li_2ZrCuO_4$ [13], via a sharp transition in the dielectric constant and the occurrence of spontaneous electric polarization vector (*P*) along some crystallographic axis.

Simple inverse Dzyaloshinsky-Moriya (IDM) or spin-current model explains the magneto-electric behavior for such existing non-collinear magnets. However, the observed *P* value is smaller than the theoretically estimated value in the quantum spin systems, which might be the result of quantum fluctuations [14, 15]. This quantum spin driven multiferroicity has created a lot of excitement at the fundamental research level. In order to understand more about the origin and the mechanism of magneto-electric coupling in the frustrated quantum spin chains, a few more varieties of potential magnetic chain systems need to be studied.

Herein, we introduce a new magnetic chain-like system $SrCuTe_2O_6$. This cubic system [16] is isostructural to $PbCuTe_2O_6$ [17]. The first and second *nn* of Cu atoms form a frustrated 3D network, whereas the third *nn* form a non-frustrated uniform chains. The magnetic properties and heat capacity of $PbCuTe_2O_6$ shows the absence of any conventional LRO down to 350 mK, despite the presence of antiferromagnetic correlations with $\theta_{CW} \approx$ -22 K. In addition, heat capacity data exhibit a broad-maximum ($T^{max}$) at 1.15 K ($T^{max}/\theta_{CW} \approx 0.05$) and a weak-kink at 0.9 K, which are rather resulted from highly frustrated network in PbCuTe$_2$O$_6$ as suggested by the structure. The hopping parameters estimated from LDA calculations also suggest the presence of a 3D frustrated network with considerable additional interactions.

Although the spin arrangement of SrCuTe$_2$O$_6$ is same as that of PbCuTe$_2$O$_6$, but the magnetic susceptibility $\chi(T)$ data surprisingly exhibit a broad maximum at 32 K. This qualitatively infers the presence of short-range interactions originating from the magnetically dominant third *nn* couplings (uniform chains). The system undergoes antiferromagnetic (AFM) ordering with the transition temperatures $T_{N1} \approx 5.5$ K and $T_{N2} \approx 4.5$ K, which further suggests the presence of non-negligible frustrated inter-chain couplings ($J_1$ and $J_2$). Our LDA calculations also confirm that SrCuTe$_2$O$_6$ has the network of uniform chains (formed by third *nn*) with various inter-chain AFM couplings. The observation of cusp-like anomaly in the dielectric constant data and the absence of spontaneous electric polarization (*P*), together point the presence of magneto-dielectric behavior below its $T_{N1}$. This new kind of uniform chains with magneto-dielectric anomalies in SrCuTe$_2$O$_6$ might open a new problem at theoretical front to understand the mechanism for magneto-dielectric behavior in 1D AFM chain systems with frustrated inter-chain couplings.

## II. EXPERIMENTAL DETAILS

The polycrystalline samples of SrCuTe$_2$O$_6$ were prepared by the solid-state reaction method. First a precursor, SrTe$_2$O$_5$, was prepared by firing the stoichiometric amounts of SrCO$_3$ and TeO$_2$ at 450º C for 48 hours in air, followed by one intermediate grinding. In second step, the obtained single phase sample of SrTe$_2$O$_5$ was mixed with CuO in a molar ratio of 1:1 and the pelletized mixture was sealed in an evacuated quartz tube and fired at 650ºC for 72 hours. Two intermediate grindings were required to get the single phase sample of SrCuTe$_2$O$_6$. The powder X-ray diffraction measurements were done at room temperature using a PANalytical X' pert PRO powder diffractometer equipped with Cu Kα radiation. The magnetic measurements were performed in the *T*-range from 2 to 300 K and in the magnetic field (*H*) range 0 - 70 kOe on a Quantum design SQUID-VSM. The heat capacity ($C_p$) measurements were done using PPMS (Quantum design, Corp.) in the *T*-range from 2 to 300 K and in the field range from 0 to 90 kOe. To ensure the better thermal contact between sample and the sample platform, Apiezon N grease was used for $C_p$ measurements. For the capacitance measurement, the pellet was cut and polished to make a thin plate of thickness about 0.3 mm and the electrical contacts were made on both sides of the pellet by attaching the thin copper wires of thickness (85 μm) with silver epoxy. Finally, the capacitance and pyroelectric current were measured using a high precision capacitance bridge (AH 2550 A) and KE617 electrometer respectively. These experimental set-up were attached to PPMS to carry out the measurements as a function of magnetic field (0-90kOe) and temperature.

## III. RESULTS AND DISCUSSION

### A) X-ray diffraction and structural features

The SrCuTe$_2$O$_6$ crystallizes in a cubic space group $P4_132$ (No. 213) [16]. Figure 1 shows the measured X-ray diffraction pattern (XRD) of polycrystalline samples. The atomic coordinates obtained as a result of Rietveld refinement, under Fullprof method, are summarized in table 1. The obtained lattice constant by the Fullprof refinement method [18] is to be $a \approx 12.463$ Å and is consistent with the previously reported value of $a \approx 12.472$ Å [16]. The selected bond angles and bond lengths between Cu atoms are provided in the table 2.

Table 1. The atomic coordinates and the occupancies obtained for SrCuTe$_2$O$_6$ after the Rietveld refinement at room temperature under the space group $P4_132$ (No. 213).

| Atom | Site | x/a | y/a | z/a | Occ. |
|---|---|---|---|---|---|
| Te | 24e | 0.33873 | 0.91845 | 0.06153 | 1 |
| Sr1 | 8c | 0.05761 | 0.05761 | 0.05761 | 1 |
| Sr2 | 4b | 0.375 | 0.625 | 0.125 | 1 |
| Cu | 12d | 0.48168 | 0.875 | 0.26832 | 1 |
| O1 | 24e | 0.72185 | 0.95322 | 0.18730 | 1 |
| O2 | 24e | 0.43581 | 0.99907 | 1.22247 | 1 |
| O3 | 24e | 0.21346 | 0.95865 | 0.13565 | 1 |

As shown in figure 2(a) and 2(b), the structure of the titled compound consists of CuO$_4$ square plates, one-end opened TeO$_3$ units, and the octahedral environments of Sr1, and Sr2. Similar to its sister compound PbCuTe$_2$O$_6$, this structure also has a similar 3D spin network of triangles formed by its first nearest neighbor *(nn)* and second *nn* couplings (see figure 2(c)). The first *nn* distance between the Cu atoms is about 4.56 Å, while second *nn* has a bond distance of about

5.52 Å. The magnetic couplings ($J_1$ and $J_2$) are mediated by O-Sr2-O path and O-Te-O paths, respectively. On the other hand, the third *nn* coupling forms the network of uniform chains via two possible paths (see figure 3(a) and table 2). One path is mediated through O-Sr2-O and another is mediated through O-O. The coupling as a result of uniform chain interaction is denoted by $J_3$.

Despite the similar structure of $SrCuTe_2O_6$ and $PbCuTe_2O_6$, there are yet some differences observed in the bond lengths and bond angles of the corresponding exchange couplings (see table 2). It is evident that the bond angles corresponding to $J_2$ ($J_3$) in Sr-system are smaller (larger) than those of Pb-system. This might suggest a relatively weaker $J_2$ (which was the strongest in Pb-system) and a stronger $J_3$ in $SrCuTe_2O_6$. In addition, Pb-system has two lone-paired elements ($Pb^{2+}$ and $Te^{4+}$) whereas the Sr-system has only one lone-pair electron ion ($Te^{4+}$). These features might be responsible for the change in bond angles and bond lengths and thus might favor a different ground state for $SrCuTe_2O_6$.

Table 2. A comparison between the bond angles and the bond lengths of exchange couplings for $PbCuTe_2O_6$ & $SrCuTe_2O_6$.

| Coupling | Paths for $SrCuTe_2O_6$ | $SrCuTe_2O_6$ | | $PbCuTe_2O_6$ | |
|---|---|---|---|---|---|
| | | Bond length (Å) | Bond angle (°) | Bond length (Å) | Bond angle (°) |
| $J_1$ | Cu-O-Sr2-O-Cu | 4.56 | O-Sr2-O ≈ 94.3 | 4.370 | O-Pb2-O ≈ 91.3 |
| $J_2$ | Cu-O-Te-O-Cu | 5.52 | O-Te-O ≈ 92.5 | 5.60 | O-Te-O ≈ 97 |
| $J_3$ | Cu-O-Sr2-O-Cu | 6.29 | O-Sr2-O ≈ 162.2 | 6.27 | O-Pb2-O ≈ 156 |
| | Cu-O-O-Cu | 6.29 (O-O ≈ 2.78 Å) | Cu-O-O ≈ 154.4 | (O-O ≈ 2.78 Å) | Cu-O-O ≈ 153.7 |

## B) Magnetic susceptibility $\chi(T)$

Magnetization, *M*, measurements were performed on the polycrystalline sample, under a magnetic field (*H*) of 10 kOe, in the temperature range 2-300 K and the magnetic susceptibility

$\chi(T)$ (= $M(T)/H$) vs $T$ plot is depicted in figure 4(a). In a sharp contrast to PbCuTe$_2$O$_6$ [17], the $\chi(T)$ data for SrCuTe$_2$O$_6$ exhibit a broad maximum at about 32 K, which signifies the presence of short-range correlations. In fact, it is quite surprising to observe such a broad maximum in a three-dimensional (3D) spin system (as anticipated from the structure). Generally the appearance of broad maximum in $\chi(T)$ data is a usual feature of the low-D (1D and 2D) spin systems.

Apart from the broad maximum at 32 K, the low temperature $\chi(T)$ data also evidence a sharp drop at 5.5 K (= $T_{N1}$), followed by a weak-kink at 4.9 K (= $T_{N2}$) measured with a field of 10 kOe. These transitions seem to be typical to an ordered antiferromagnetic system. The fitted $\chi(T)$ data to the Curie-Weiss law ($\chi_o + C/(T-\theta_{CW})$) in the $T$-range from 70 K to 300 K (not shown) yield temperature independent susceptibility ($\chi_o$), the Curie-constant (C= $Ng^2\mu_B^2S(S+1)/3k_B$), and the Curie-Weiss temperature ($\theta_{CW}$) to be (1.30 ± 0.05) x 10$^{-4}$ cm$^3$/mol-Cu, (0.42±0.01) cm$^3$ K/mol-Cu, and -(42±2) K, respectively. Here $N_A$, $g$, $\mu_B$, and k$_B$ are the Avogadro number, the Lande-g factor, the Bohr magneton, and the Boltzmann constant, respectively. The diamagnetic susceptibility ($\chi_{dia}$) of SrCuTe$_2$O$_6$ was calculated to be -1.52x10$^{-4}$ cm$^3$/mol Cu from $\chi_{dia}$ of their ions (Sr$^{2+}$, Cu$^{2+}$, and (TeO$_3$)$^{2-}$) [19]. The estimated Van-Vleck paramagnetic susceptibility $\chi_{VV}$ (=$\chi_o$-$\chi_{dia}$) gives the value of 2.2x 10$^{-5}$ cm$^3$/mol, which is in good agreement with other cuprates [20-22]. A negative $\theta_{CW}$ value suggests that spin interactions are of antiferromagnetic nature. The estimated effective moment ($\mu_{eff}$) value of Cu$^{2+}$ is 1.83 $\mu_B$, which is comparable to that of $S$=1/2 (1.73 $\mu_B$) moment.

In order to understand the ground state behavior of SrCuTe$_2$O$_6$ and the cause which drives the system into a different ground state other than that of PbCuTe$_2$O$_6$ is of paramount importance to understand the behavior of exchange couplings in this system. We looked at the different models of isolated $S$=1/2 triangular and $S$=1/2 hyper-Kagome lattices (first and second *nn* couplings as suggested by the structure) but none of the above model produce broad maximum in their magnetic susceptibilities data. With the suspicion of third *nn* coupling being the dominant one, we looked at the uniform chain model, which has a broad maximum in the magnetic susceptibility data. To support this reasoning, we also performed the *LDA* band structure calculations (explained in a latter section) to know the relative strengths of exchange couplings. From the LDA calculations, we found that the major interaction coupling is the third *nn* which essentially forms a network of uniform chains, propagating along all the three crystallographic (*x*-, *y*-, and *z*-) axes, as shown in figure 2(d). Having identified the third *nn* coupling being the leading exchange interaction in our system, we finally analyzed our magnetic data in the framework of *S*=1/2 uniform spin chain model [23]. A successive appearance of anomalies below 5.5 K is an indication of the involvement of the inter-chain couplings. So in the view of this fact the data were finally fitted to the coupled *S*=1/2 uniform chain expression (1) down to 12 K, taking inter-chain couplings into consideration.

$$\chi = \chi_0 + \frac{\chi(J_3, T)}{1 + \frac{zJ'}{Ng^2\mu_B^2}\chi(J_3, T)} \tag{1}$$

The obtained value of $\chi_0$ and Lande g-factor are consistent with the previous values obtained from the Curie-Weiss fit. The values of uniform intra-chain coupling $J_3/k_B$ and the total strength of inter-chain coupling *ZJ'* are found to be (43±2) K and (13±1) K, respectively. The obtained value $J_3/k_B$ also matches with the expected value for 1D chain with a broad maximum 32 K ( ≈ 0.640851 $J_3/k_B$). To calculate the inter-chain coupling (*ZJ'*) just by exploiting the values of $T_N$ and $J_{chain}$ using the following expression (2) [23]

$$zJ' = \frac{T_N}{0.2333\sqrt{\ln\left(\frac{2.6J_3}{T_N}\right) + \frac{1}{2}\ln\left(\ln\left(\frac{2.6J_3}{T_N}\right)\right)}} \tag{2}$$

The estimated value of *ZJ'~12* comes very close to the previously obtained value from the expression (1).

**C) Heat capacity ($C_p$) in zero field**

The *T*-dependent heat capacity $C_p$ of the polycrystalline SrCuTe$_2$O$_6$ sample in zero-field is shown in figure 4(b). Being a magnetic insulator, SrCuTe$_2$O$_6$ has two contributions to $C_p$ namely; lattice and magnetic. Due to the lack of availability of non-magnetic analog of SrCuTe$_2$O$_6$, we used the Debye model [24] to extract the lattice contribution of the $C_p$ data. We found that a linear combination of two Debye integrals results the best fit to the $C_p$ data. The data are fitted successfully, in the *T*-range from 40 K to 130 K, using the following equation mentioned in the text.

$$C_p = 9rNk_B \sum_{i=1,2} C_i \left(\frac{T}{\theta_D^i}\right)^3 \int_0^{x_D^i} \frac{x^4 e^x}{(e^x-1)^2} dx \tag{3}$$

Here *r* is the number of atoms per formula unit, $\theta_D$ is a Debye temperature. The fitting yields

$C_1 \approx (0.3 \pm 0.05)$, $\theta_{D1} \approx (171 \pm 5)$ K, $C_2 \approx (0.46 \pm 0.05)$, and $\theta_{D2} \approx (562 \pm 10)$ K. The fitted curve was then extrapolated down to 2 K and subtracted it from the measured $C_p(T)$ data. As a consequence, the obtained magnetic heat capacity $C_m(T)$ is shown in the inset (ii) of figure 4(b). Similar to $\chi(T)$ data, two sharp and distinctive anomalies are observed, as can also be seen on the plot of $C_p/T^2$ vs $T$ (inset (i) of figure 4(b)). The data follow $T^3$ behavior at low-$T$, expected for AFM ordered spin systems, which are contrary to the $T^2$ behavior observed in $PbCuTe_2O_6$.

We observed a broad maximum ($T^{max}$) at 15 K in the $C_m(T)/T$ data and it is expectedly lower than the value of broad maximum at 32 K in $\chi$ data. This feature has already been seen in several low-$D$ spin systems [23, 25]. The observed values of $T^{max}$ and $(C_m/T)^{max}$ also provides a way to estimate the exchange energy couplings in the system using theoretical equations given below for 1D model. These values are nearly in agreement with the value estimated from $\chi$ data.

$$(T_{max})^{C_m/T} = 0.307 \left(\frac{J_3}{k_B}\right) \tag{4}$$

$$(C_m/T)^{max} = 0.8973651 \frac{Nk_B^2}{J_3} \tag{5}$$

The obtained values are found to be $J_3/k_B \approx 43$ K and 38 K, respectively. A small disagreement is observed in the above estimated values might be due to the error in estimating the magnetic heat capacity.

The entropy change $S_m$ calculated from the magnetic heat capacity is found to be 5.5 J/mol K, which is very close to the expected value $Rln2$ for $S = 1/2$ systems (see inset (ii) of figure 4(b)). The $S_m$ value at transition $T_{N1}$ is found to be $\approx 0.6$ J/mol K, which is only about 10% of the total entropy and the rest of the entropy is recovered in the paramagnetic region up to the $\theta_{CW}$ temperatures (well above $T_N$). This is also an indication for the presence of short range correlations in the paramagnetic region.

**D) Dielectric constant $\varepsilon'(T)$ in zero field**

Figure 4(c) displays the temperature dependence of dielectric constant, $\varepsilon'(T)$, of $SrCuTe_2O_6$ measured in zero field. As $T$ decreases, the $\varepsilon'(T)$ increases and shows a cusp-like anomaly at 5.5 K, followed by another anomaly at 4.5 K, respectively. This is in the good agreement with the magnetic anomalies previously seen in $\chi(T)$ and $C_p(T)$ data at zero field. The dielectric loss ($\Delta Tan\delta$) is found to be less than 0.1 %, which is almost negligible. The simultaneous appearance of these transitions in $\varepsilon'(T)$, $C_p(T)$ and $M(T)$ data suggest the presence of magneto-dielectric behavior which is usually a common feature for magneto-dielectric materials. For example $CuTeO_3$, $CuSeO_3$, *etc.* [26] also show a similar cusp-like behavior but it is rather different from the λ-like peak feature noticed for other multiferroics, where the spontaneous ferroelectricity *P* appears below the transition temperature. The absence of pyrocurrent below this transition probably hints the absence of spontaneous ferroelectricity in this system at zero field.

*E) $M(T)$, $C_p(T)$, and $\varepsilon'(T)$ in magnetic fields*

To know more about the nature of low-*T* AFM and dielectric transitions, we measured $M(T)$, $C_p(T)$ and $\varepsilon'(T)$ in different magnetic fields ranging from 10-90 kOe. The $\chi(T)$ data in different magnetic fields are shown in figure 5(a). The $\chi(T)$ data show a field dependent behavior below about 60 kOe and $T_{N1}$ & $T_{N2}$ gradually move to the lower temperatures on increasing the magnetic field. However, $T_{N2}$ shifts rapidly towards the low-*T* side with the applied fields and finally suppressed by the magnetic field of about 40 kOe (called as $H_{C1}$). Similarly, the other anomaly $T_{N1}$ is also suppressed by 60 kOe (called as $H_{C2}$). To exaggerate the anomalies at low temperatures, the derivative plot of $\chi$ with respect to *T* (*i.e.*, $d\chi/dT$) is shown in the inset of figure 5(a). In this plot, the 30 kOe data has two peaks corresponding to $T_{N1}$ and $T_{N2}$, however, at 50 kOe the peak corresponding to $T_{N1}$ remains and $T_{N2}$ suppresses completely. Finally at 70kOe the peak associated with $T_{N1}$ also disappears and the data leftover with a small dip at 5.3 K.

The similar behavior is also observed in $C_p(T)$ and $\varepsilon'(T)$ data measured with different fields, as shown in figure 5(b) and 5(c), respectively. Likewise to the magnetic data, the $C_p/T$ data at 20 kOe also show two distinctive peaks corresponding to $T_{N1}$ and $T_{N2}$ (see inset of figure 5(b)). Having suppressed $T_{N2}$ by a field of 40 kOe, the data of $C_p/T$ at 50 kOe left only with a single sharp peak. When the field was increased up to 70 kOe, the magnitude of the peak at $T_{N1}$ was suddenly dropped and the peak became broaden. This very feature points a change in the antiferromagnetic nature of $T_{N1}$ in this region and on further increasing the field, above 70 kOe, no much change in the shift as well as the shape of the peak was observed. Here we note that, this robust peak survives in $C_p$ for *H>70* kOe, whereas such a peak was not found in the magnetic data. We suspect that the peak might be hidden under the Curie behavior in the magnetic data (see figure 5(a)). In order to reassure $\chi(T)$ and $C_p(T)$ findings, we also performed

$\varepsilon'(T)$ in a field range 0- 90 kOe. The results of $\varepsilon'(T)$ nearly matches with $C_p$ and $\chi$ data. Likewise, the shift in the peak position and a difference in the shape of the anomalies with the applied magnetic fields at different regions are shown in figure 5(c).

*F) M(H), $\varepsilon'(H)$ at different temperatures*

To unveil this field-induced magnetic behavior, we measured *M(H)* isotherms at different temperatures from 2 to 5 K (see figure 6(a)). At 2 K, *M(H)* shows two jumps at the fields about 40 kOe (SF1) and 53 kOe (SF2), respectively and suggest that these field-induced transitions are due to the reorientations of the spins in the ordered state, as a result they seem to be spin-flop transitions. The SF1 decreases as *T* increases and finally disappeared at *T* above 4 K, while the SF2 increases first and then again decreases with *T*. This behavior can be clearly seen in d*M*/d*H* vs. *H* plot in the inset of figure 6(a). The field-induced spin-flop transitions SF1 and SF2 are absent in the *M(H)* data at temperatures above $T_{N1}$. The value of fields at which SF1 and SF2 observed are nearly equal to the value of the critical fields $H_{C1}$ and $H_{C2}$, (fields at which the $T_{N2}$ and $T_{N1}$ are suppressed in the $\chi(T)$), respectively. This again suggests that the observed field-induced transitions ($T_{N1}$ and $T_{N2}$) are rather related with these spin-flop transitions. The similar behavior is also observed in $\varepsilon'(H)$ and $d\varepsilon'/dH$ data at different temperatures as shown in figure 6(b) and its inset .

**G) Magnetic phase diagram**

From magnetization (*M*), heat capacity ($C_P$), dielectric constant ($\varepsilon'$) data and their derivatives with respect to *H* and *T*'s, a magnetic phase-diagram is built, as shown in figure 7. The magnetic phase-diagram separates the boundary between paramagnetic (PM) and antiferromagnetic (AFM) regions. Also, it has been identified that the AFM region is comprised of different phases AFM-1, AFM-2 and AFM-3 which are found to be tuned by the external magnetic field. Detailed neutron diffraction study is needed to understand the spin orientations of these local AFM regions.

**H) Electronic structure calculations**

In order to understand the magnetic behavior and the basic electronic structure, we have carried out density functional theory calculations using Vienna *ab-initio* simulation package (VASP) code [27, 28] within projector augmented-wave (PAW) method [29, 30]. Exchange and correlation effects are treated using the local density approximation (LDA). The kinetic energy cut off of the plane wave basis was chosen to be 600 eV. Brillouin-Zone integration have been performed using 4x4x4 *k*-mesh.

Figure 8(a) display the non-spin polarized band dispersion of SrCuTe$_2$O$_6$ along various high symmetry directions of the Brillouin zone corresponding to the cubic lattice. The most important feature of the band structure is the isolated manifold of twelve bands near the Fermi level ($E_F$) which arises from the twelve Cu atoms in the unit cell. These bands are predominantly of Cu $d_{x2-y2}$ character in the local frame of reference where the Cu atom is at the square planar environment of O atoms. The crystal field splitting corresponding to a square planar environment has been shown in the inset of figure 8(a). Since Cu is in $d^9$ configuration, these isolated bands are half filled and separated from the other Cu *d* bands by a gap of 1.1 eV (see figure 8(a)), and hence these bands are responsible for the low energy physics of the material.

To extract a low energy model Hamiltonian, we construct the Wannier function for the $d_{x2-y2}$ like bands, using the VASP2WANNIER and the WANNIER90 codes [31].

Figure 8(b) displays the superimposed Wannier-interpolated bands on the LDA bands and the agreement is quite remarkable. The various hopping interactions ($t_n$), obtained with this method are shown in table 3. Since the only relevant orbital is $d_{x2-y2}$ which is half filled, Cu at different sites can only interact via the super-exchange mechanism. The super-exchange interactions between the Cu ions at different sites, have been estimated using the relation *J = 4t$^2$/U$_{eff}$*, where $U_{eff}$ is the effective onsite Coulomb interaction. Since a constrained DFT calculation by Anisimov *et al* [32] for CaCuO$_2$, gives U$_{eff}$ = 6.5 eV for the Cu ions where Cu is in the same charge state 2+ as in the present system, we choose this value for the estimation of *J*. Our estimated *J* (see table 3) clearly reveals that the third *nn* exchange interaction ($J_3$) is the most dominant one. To understand why $J_3$ is the most significant interaction in this system, we analyze the interesting crystal geometry of this system. The third *nn* Cu atoms interact via Cu-O-O-Cu path with O-O bond (see figure 3(a)) and hence the third *nn* Cu form a one dimensional chain. The Cu $d_{x2-y2}$ strongly hybridize with O $p_x$ orbital via bonding O-O which mediate the Cu ions are close to each other. As a consequence, Cu $d_{x2-y2}$ – Cu $d_{x2-y2}$ hopping has become the most significant for the third *nn*, as also shown in Wannier plot (see figure 9). We also found that the next dominant interaction is the second *nn* ($J_2$), which is relatively small ($\approx 0.1 J_3$) and also frustrated, hence this can only play a significant role at low temperatures. As shown in table 3, the remaining interactions are very small and do not play any decisive role in the magnetism of this system. Since $J_2$ is quite small and $(1/10)^{th}$ of $J_3$, we can consider the SrCuTe$_2$O$_6$ as a *1D* uniform chain with a very small frustrated inter-chain interactions.

Table 3. The hopping integrals and the exchange interaction between the Cu ions.

| Hopping | Cu-Cu distance (Å) | Hopping parameters (meV) | $J_i/J_3 = (t_i/t_3)^2$ | Exchange interactions |
|---|---|---|---|---|
| $t_1$ | 4.55 | 13.99 | 0.03 | 1.36 |
| $t_2$ | 5.52 | 26.54 | 0.11 | 5.03 |
| $t_3$ | 6.29 | 79.92 | 1.00 | 45.60 |

From the values obtained from LDA calculations, one can estimate the total strength of inter-chain interactions $zJ' = (z_1J_2 + z_2J_2) \approx 0.5\ J_3$. According to the uniform chain model with unfrustrated inter-chain couplings (Eq. 2), this value supposed to have a $T_{N1}$ value of about 17, which is, in fact, very much larger than that of experimentally observed value of 5.5 K ($\approx T_{N1}$). This disagreement suggests that the inter-chain interactions in SrCuTe$_2$O$_6$ are in the frustrated nature, so it results in a smaller $T_{N1}$ value. Such kind of differences are also observed in several earlier found spin chains with frustrated inter-chain couplings: Ca$_2$CuO$_3$ [33, 34], Sr$_2$Cu(PO$_4$)$_2$ [35, 36], K$_2$CuP$_2$O$_7$ [37] *etc*.

However, we did not observe any electric polarization $\leq T_{N1}$ in this material unlike to the other intra-chain frustrated, multiferroic chain materials [5-12]. This might be due to the lack of sufficient strong inter-chain couplings ($J_2/J_3$ is about 0.1) in this material. A further theoretical study is required to know the origin of magneto-dielectric behavior in this material.

## IV. CONCLUSION

We have studied the magnetic properties of a new chain material SrCuTe$_2$O$_6$ via magnetic, specific heat, and dielectric constant measurements and electronic structure calculations. However, the crystal structure of SrCuTe$_2$O$_6$ is similar to its sister compound PbCuTe$_2$O$_6$, but due to the dominant third *nn* exchange coupling the ground state exhibits the characteristic features of one-dimensional magnetism. The magnetic data analysis well corroborated by LDA band structure calculations also suggest the presence of uniform chains with non-negligible frustrated inter-chain couplings ($J_1$ and $J_2$), which might lead to magneto-dielectric anomalies observed at low-*T* (5.5 K and 4.5 K). Magnetic field induced phases are also observed in antiferromagnetically ordered region. The magnetic phase diagram built on behalf of the magnetization, heat capacity, and dielectric constant experiments evidences the presence of different AFM regions. The determination of magnetic structures of those phases by neutron diffraction will be the scope of future study to explore the associated mechanism of magneto-dielectric effect in this new type of *S=1/2* chain material with frustrated inter-chain interactions.

**Acknowledgments:** B.K. thanks DST INSPIRE faculty award-2014 scheme. F.C.C. acknowledges the support from the National Science council of Taiwan under project number NSC-102-2119-M-002-004. A.V.M. thanks the Department of Science and Technology, Government of India for financial support. R. Kumar acknowledges IRCC IITB grant for financial support. The work at SNU was supported by the National Creative Research Initiative (2010-0018300). We thank Patrick Lee for fruitful discussions.

Electronic address:* khkim@snu.co.kr, + fcchou@ntu.edu.tw

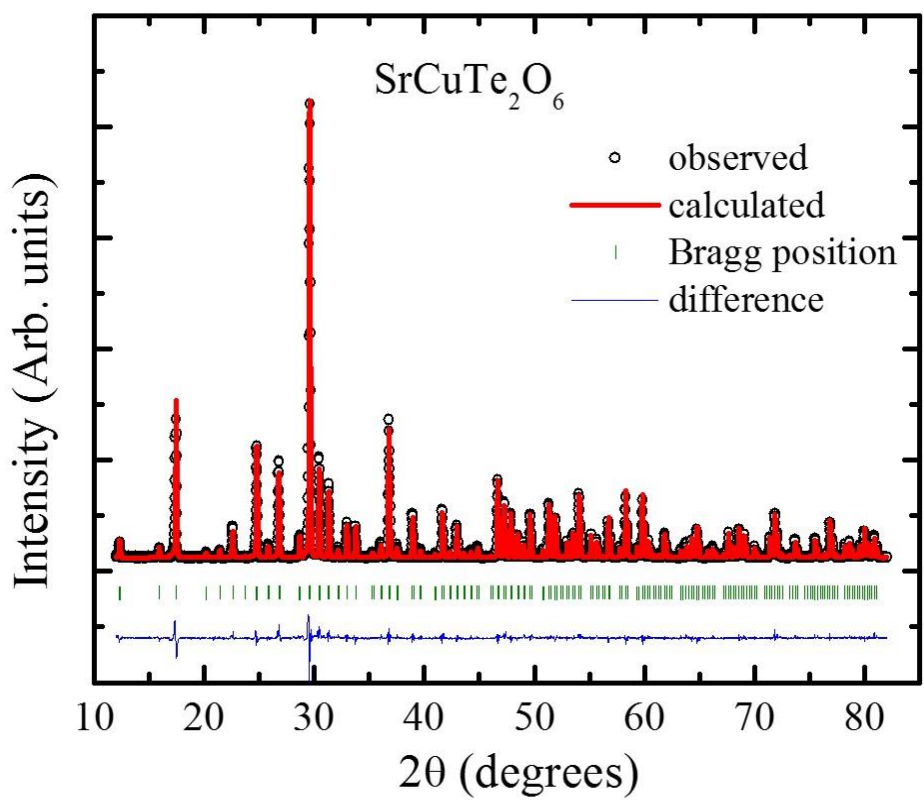

Figure 1 (color online). The Rietveld refinement of the powder x-ray diffraction pattern collected for $SrCuTe_2O_6$ at room temperature. The open circles (black) indicate the experimentally collected data, while the Rietveld refinement fit is shown with a red solid line. The Bragg positions and the difference curve are denoted by vertical green bars and blue line, respectively. The residual parameters for the Rietveld refinement are $R_p \approx 4.388\ \%$, $R_{WP} \approx 6.469\ \%$, $R_{exp} \approx 2.369\%$, $R_{Bragg} \approx 3.235\ \%$, and goodness of fit (GOF) = $R_{WP}/R_p \approx 1.474$, respectively.

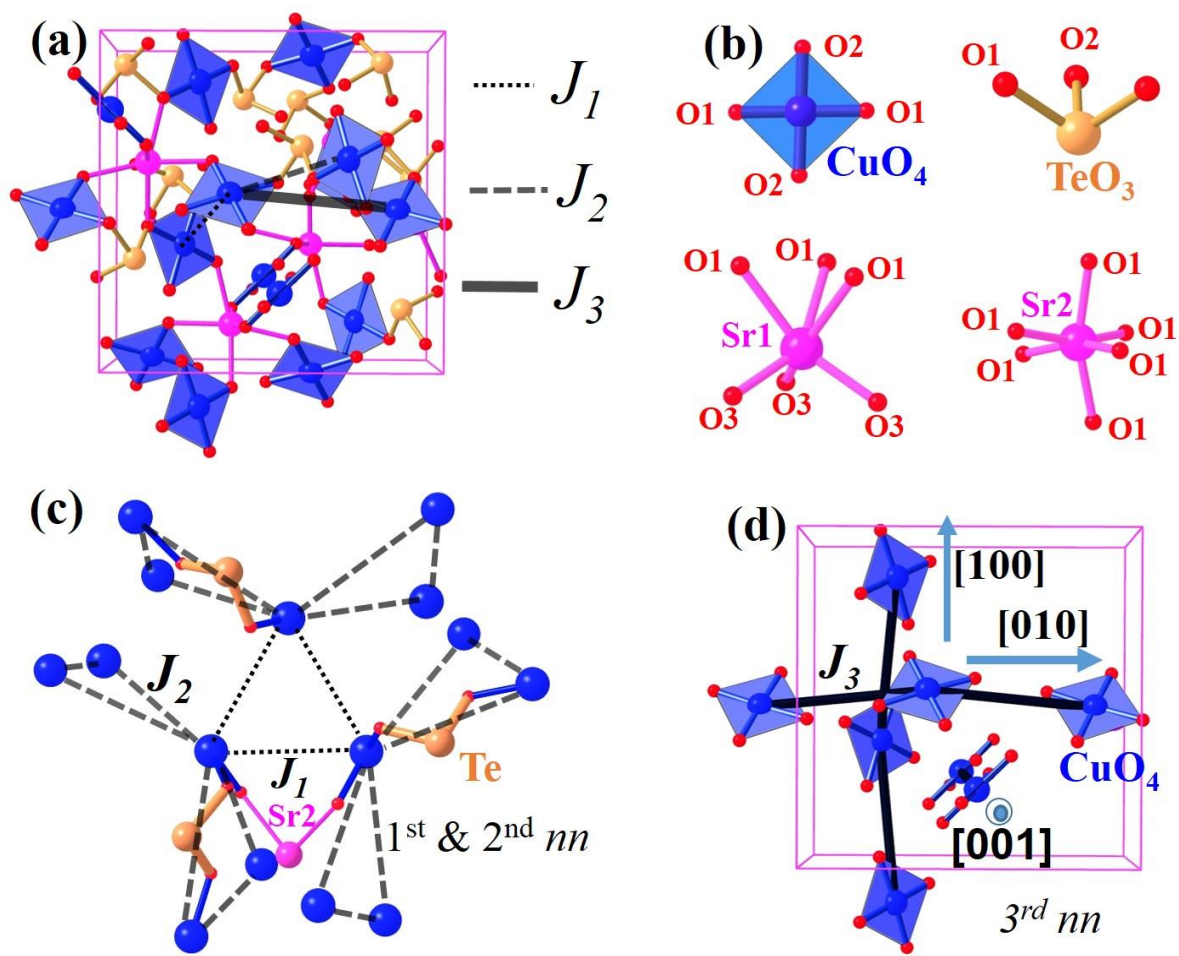

Figure 2 (color online). (a) Crystal structure of SrCuTe$_2$O$_6$.[16LWulf1997] (b) The environments of CuO$_4$, TeO$_3$, Sr1O$_6$, and Sr2O$_6$ units. (c) Formation of the *3D* network of corner-sharing triangles by the first *nn* and second *nn* coupled Cu-atoms. (d) Formation of uniform chains by the third *nn* coupling, passing along all the crystallographic (*a*, *b*, and *c*) directions.

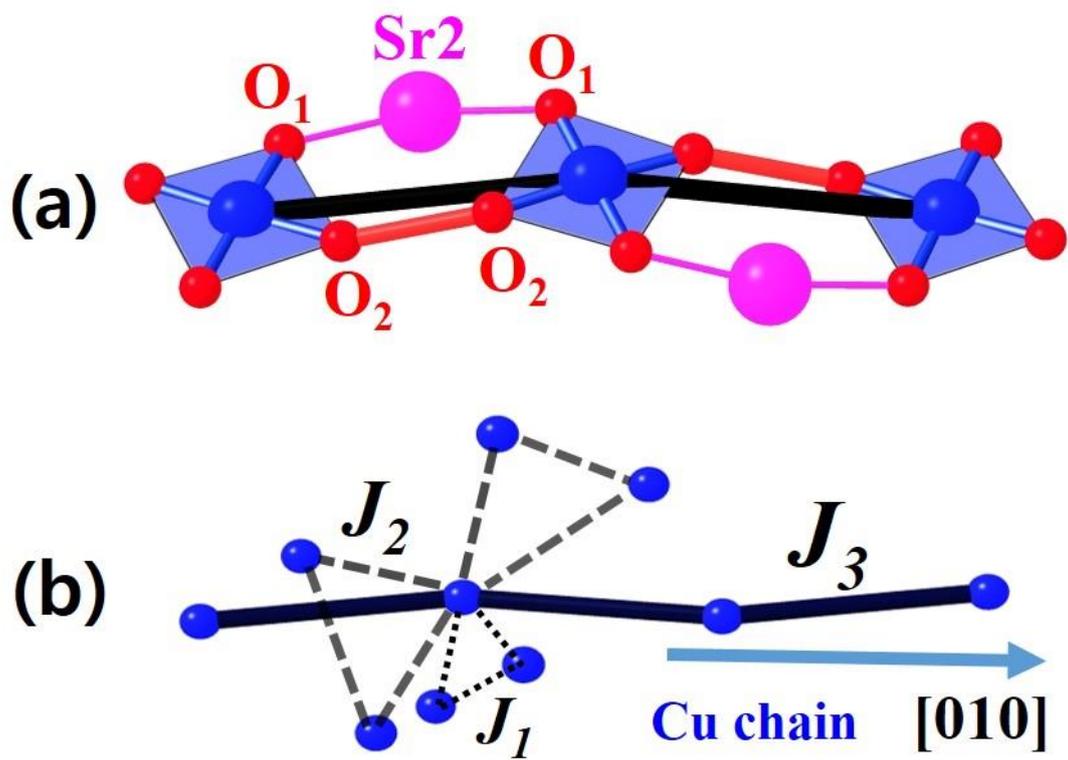

Figure 3 (color online). (a) The details of the uniform spin chain formed by the third *nn* coupling ($J_3$). (b) The uniform spin chain with its associated frustrated inter-chain couplings ($J_1$ and $J_2$).

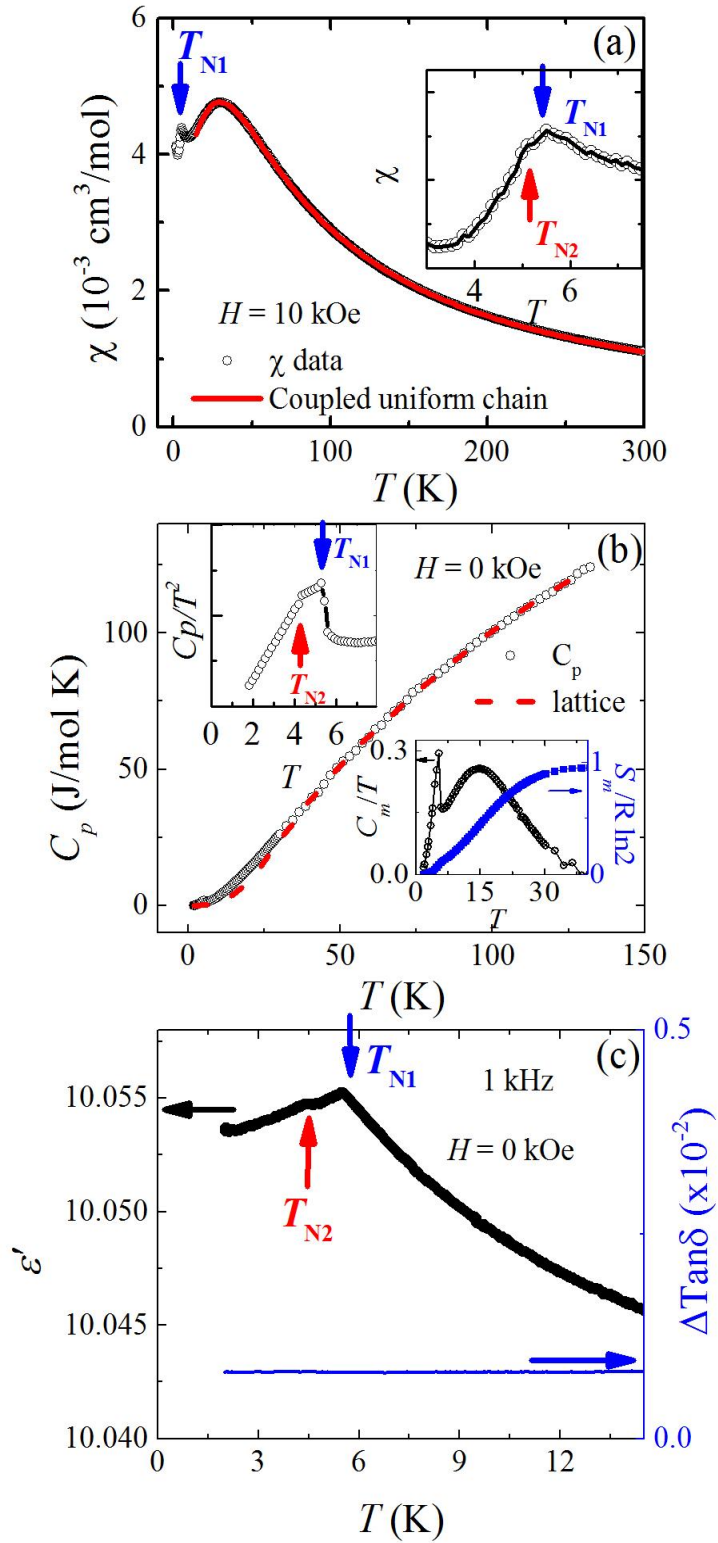

Figure 4 (color online). (a) Magnetic susceptibility $\chi$ versus temperature $T$ with a fit to a coupled $S=1/2$ uniform chain model (red solid line). Inset shows the appearance of two transitions $T_{N1}$ and $T_{N2}$ at the low temperatures. (b) Zero field $C_p$ with a fit to the lattice contribution. Inset (i) shows the plot of $C_p/T^2$ versus $T$. Inset (ii) shows the plot of magnetic heat capacity $C_m/T$ (left-axis) and the normalized magnetic entropy change $S_m/Rln2$ (right-axis) versus $T$. (c) The plot of zero field dielectric constant (left-axis) and dissipation (right-axis) versus $T$. The two transition temperatures $T_{N1}$ and $T_{N2}$ are indicated by down and up arrow marks, respectively.

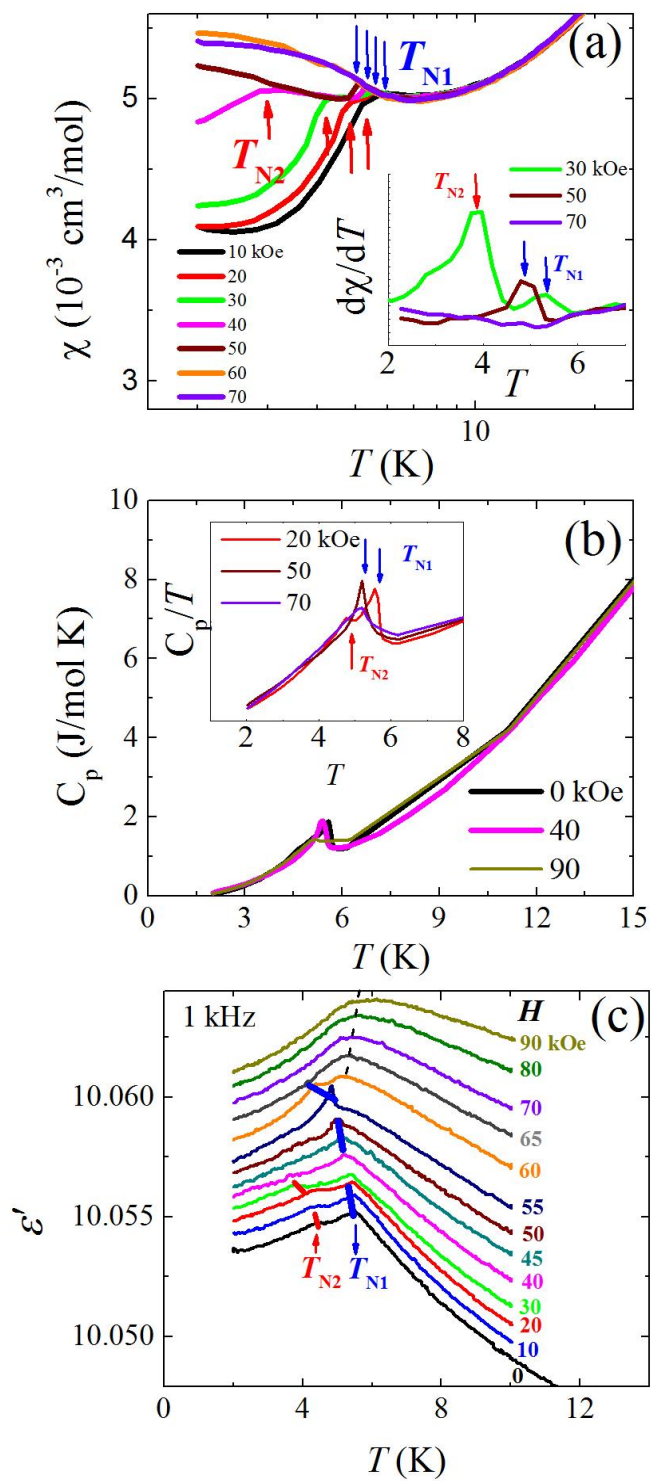

Figure 5 (color online). (a) $\chi(T)$ data in different magnetic fields from 10 kOe to 70 kOe. Inset of (a) shows the plot $d\chi/dT$ versus $T$. (b) The main (inset) plot shows $C_p$ ($C_p/T$) versus $T$ for different fields up to 90 kOe. (c) The plot of dielectric constant ($\varepsilon'$) versus $T$ in different $H$'s up to 90 kOe. The two transitions $T_{N1}$ and $T_{N2}$ are indicated by down and up arrows, respectively.

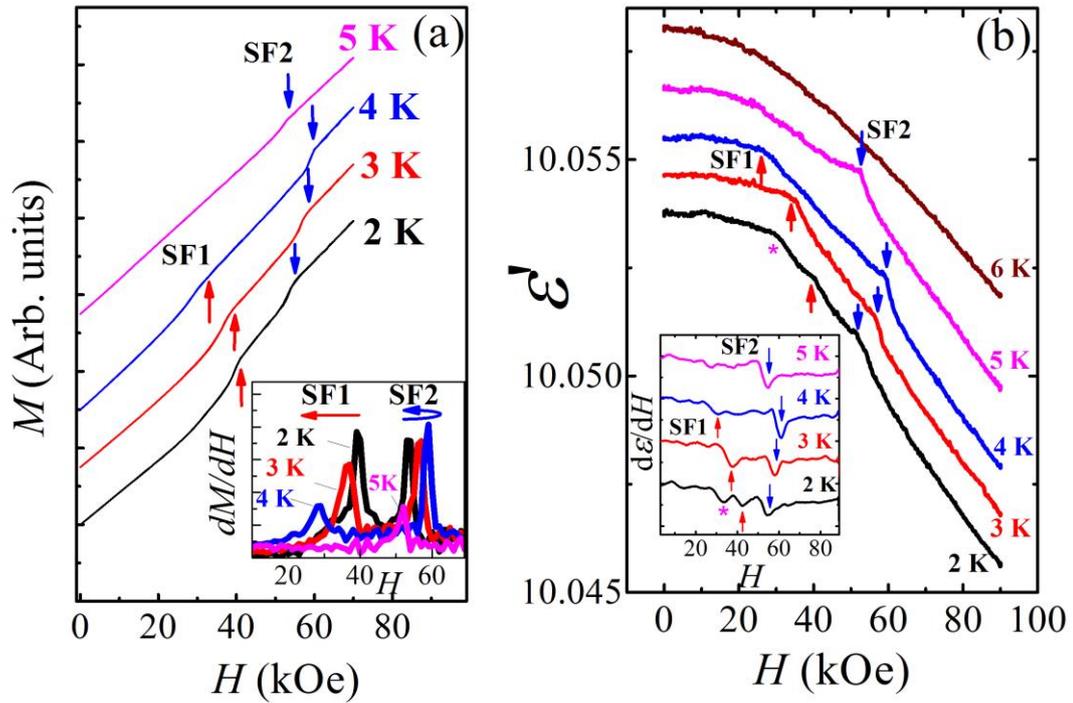

Figure 6 (color online). (a) The magnetization ($M$) versus $H$ plot at different temperatures and the inset of (a) shows the derivative plot of magnetization ($dM/dH$) versus $H$. (b) shows the dielectric constant ($\varepsilon'$) versus $H$ plot for various $T$'s from 2 to 6 K and the inset represents the $d\varepsilon'/dH$ versus $H$ plot. The vertical up and down arrow marks indicates the change in their slopes marked by SF1 and SF2, respectively. The (*) symbol indicates a change in the slope of $\varepsilon'$ data, but the corresponding change is not seen in the magnetic data.

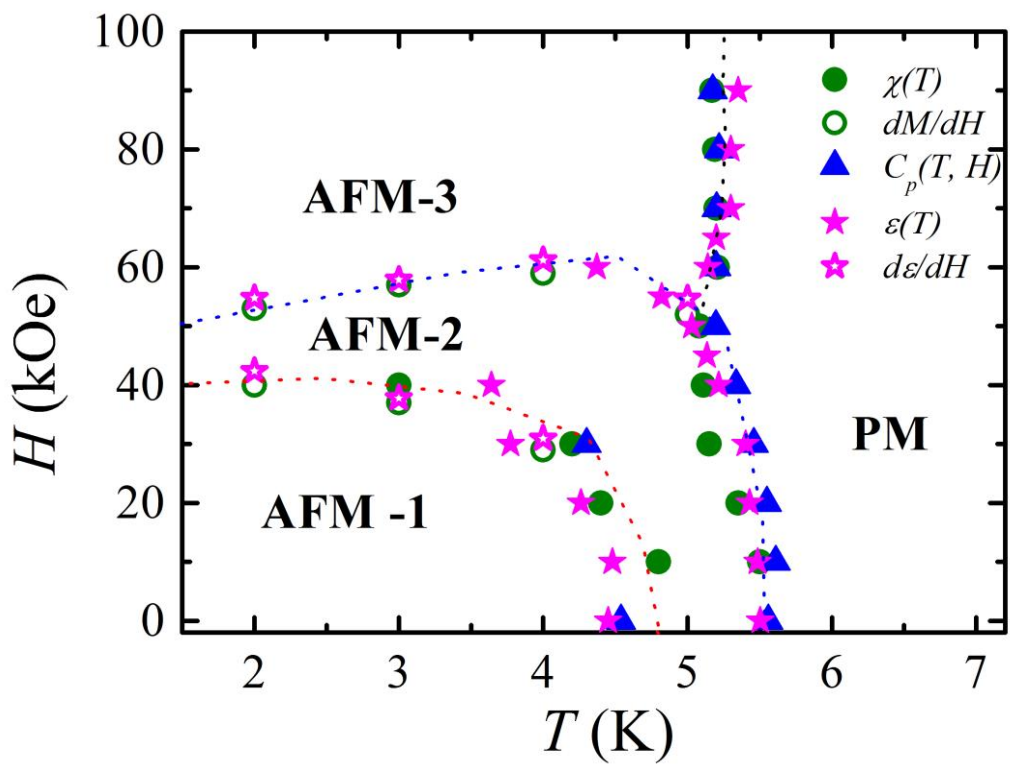

Figure 7 (color online). Evolution of different antiferromagnetic (AFM) regions (AFM-1, AFM-2 & AFM-3) and their separation from paramagnetic (PM) phase is depicted as a function of magnetic field and temperature.

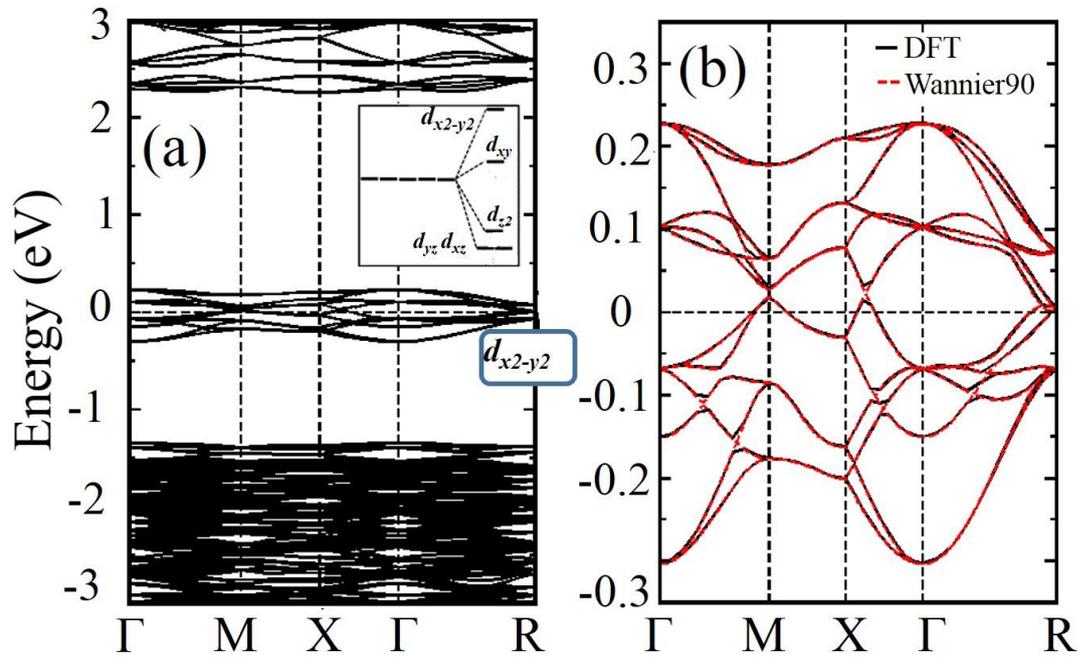

Figure 8 (color online). (a) Non-spin polarized band dispersion along various high symmetry directions. Inset shows the crystal field splitting corresponding to square planar environment. (b) Superimposed Wannier-interpolated bands on the LDA bands.

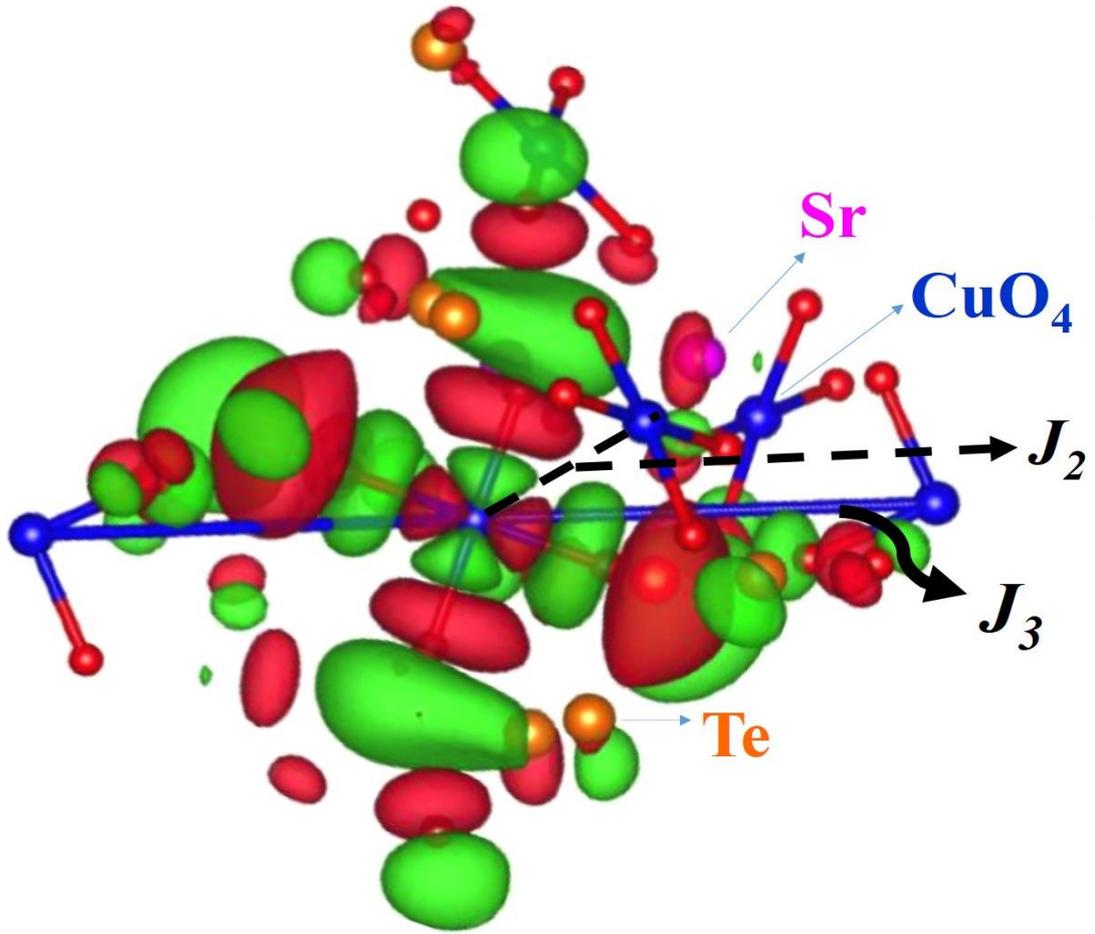

Figure 9 (color online). Wannier function plot of effective Cu $d_{x^2-y^2}$ orbitals.